\documentclass[12pt, a4paper, notitlepage, openany]{article}

\usepackage[T1]{fontenc}

\usepackage{amsmath}
\usepackage{graphicx}
\usepackage{tabularx}

\newcommand{\qq}[1]{``{#1}''}
\newcommand{\url}[1]{\texttt{#1}}

\newcommand{\cf}{\textit{cf.}}
\newcommand{\eg}{\textit{e.g.}}
\newcommand{\etal}{\textit{et al.}}

\newcommand{\ie}{\textit{i.e.}}

\title{Scope of physics-based simulation artefacts}
\author{Martin Thomas Horsch,\footnote{martin.thomas.horsch@nmbu.no}~ Fadi Al Machot, and Jadran Vrabec}
\date{}

\begin{document}

\maketitle

\begin{abstract}
Data and metadata documentation requirements for explain\-able-AI-ready (XAIR) models and data in physics-based simulation technology are discussed by analysing different perspectives from the literature on two core aspects: First, the scope of the simulation; this category is taken to include subject matter, the objective with which the simulation is conducted, and the object of reference, \ie, the simulated physical system or process. Second, the artefacts that need to be documented in order to make data and models XAIR, and modelling and simulation workflows explainable; two CEN workshop agreements, MODA and ModGra, are compared for this purpose. As a result, minimum requirements for an ontologization of the scope of simulation artefacts are formulated, and the object-objective abstractness diagram is proposed as a tool for visualizing the landscape of use cases for physics-based simulation.
\end{abstract}

\section{Introduction}

It is the overarching theme of the DCLXVI 2024 International Workshop, Kaiserslautern, to explore the conceptual landscape for the next generation of good practice recommendations in data management, beyond the FAIR principles~\cite{CGHHJLMPW18,Guizzardi20,WDA3B3DB3CCDDEEFG5HTHK3LMMPPRRVS6TV3W3ZM16}, through which modelling, simulation, and the associated data and artefacts will systematically become ready for explainable modelling and simulation workflows~\cite{ACCHSSW24}. Only such workflows can meet industrial, societal, and regulatory requirements for transparency in high-risk AI systems, and by extension in all AI systems used in environments where occasional mistakes are seen as undesirable, even where they may not be classified as high-risk technically. This has become a community concern in research data management, with a series of ideas being proposed under various labels that suggest an extension or revision of FAIR~\cite{Bashir25}. These include FAIR+~\cite{CD21}, which has a focus on eliminating dark data~\cite{CC22,Schembera21}, FAIR 2.0~\cite{VSMKKPB24}, which targets improved semantic interoperability, and the design specifications for FAIR digital objects~\cite{ABBHIJLPQS4VWWZ23b}, \eg, on machine-interpretability and machine-actionability~\cite{WIBAW22}. The Knowledge Graph Alliance's working group on explainable-AI-ready data and metadata principles\footnote{URL: \url{https://www.kg-alliance.org/kga-wg-xai-24-4/}.} has adopted the abbreviation XAIR, for \textit{explainable-AI-ready}~\cite{HSP23}, for its line of work toward a community-driven dialogue on good practice with an orientation toward documenting the knowledge status of models as well as data and artefacts related to modelling~\cite{HCTCDKSSTSKAS24}.

The present work contributes to this effort by \qq{surveying the landscape of multiple core concepts} through \qq{an analysis of how and whether different definitions of concepts relevant to XAI-readiness can be combined with each other}~\cite{ACCHSSW24}. The two high-level concepts considered here are \textit{scope} and \textit{simulation artefact}. Scope is whatever a simulation is for, about, or applicable to, or what question it is answering. Simulation artefacts are the entities that we must make XAIR so that explainable simulation workflows become well-documented. For this purpose, we will proceed as follows: In Section~\ref{sec:scope}, scope is considered from multiple perspectives: The simulation objective following Dur\'an~\cite{Duran23}, the epistemic status of the simulation's object of reference, and subject matter following Yablo~\cite{Yablo14}. The latter is compared to the more widespread, but much less expressive approach of documenting topics as \qq{bags of words.} Section~\ref{sec:simulation-artefacts} compares simulation artefacts, \ie, documentable entities, from two European standardization documents for physics-based simulation: MODA~\cite{Cenelec18} and ModGra~\cite{Cenelec22}. In Section~\ref{sec:landscape}, these two threads are combined; as a result, we propose the \textit{object-objective abstractness diagram} as a technique for visualizing where a use case is positioned within the broad landscape of what can be done with physics-based simulation, and deduce metadata standardization requirements for the documentation of simulation artefacts and their scope. In Section~\ref{sec:conclusion}, we conclude by appealing to all stakeholders to support the Knowledge Graph Alliance in its efforts.

\section{Dimensions of scope}
\label{sec:scope}

\subsection{Objective and object}
\label{subsec:objective-object}

Nobody does a simulation by accident. A simulation is an intentional activity carried out consciously and with a purpose by its agent, using models with computational representations and solvers designed for such a purpose. Therefore, one of the dimensions of the scope of a simulation is \textit{what the agent is simulating for}, in other words, the objective of the simulation. This is closely related to the status of \textit{what the agent is simulating}, namely, the object of the simulation.

Humans can follow any kind of aims and, in the process of this, attempt to make use of any kind of tools. So in effect, anybody with access to the required hardware and software can run a simulation for any purpose; \eg, in gaming, a physically realistic model environment is put to a recreational use, and other cultural interpretations can be made of it, such as that of simulation as theatre~\cite{Laurel91,RG11}.
Attempts at prescriptively cataloguing what can be possible objectives of a simulation will therefore fail -- anything that humans aim at, they can also aim at using a simulation. Instead, let us here restrict ourselves to addressing computer simulation insofar as it is used an epistemic technology, designed to support \qq{the broader category of practices that relate in a direct manner to the acquisition, retention, use and creation of knowledge}~\cite{Alvarado23}.
In such a context,\footnote{This includes the use of simulations in teaching, \cf~Roberts and Greene~\cite{RG11}.} its use can be either \textit{scientific}, operating over an open epistemic space, where novel kinds of knowledge can be uncovered, or \textit{technical}, if the epistemic space is closed, \ie, not open to renegotiation, such as when we are only interested in figuring out the value of some property without second-guessing the conceptual scheme within which the simulated object is understood to have such a property~\cite{Tulatz18}. The above does not mean that there cannot be, at the same time, other objectives related to the simulation, beyond the technical or scientific and possibly unrelated to its use as an epistemic technology.

The objective of a simulation and the status of the simulated object are two independent dimensions of scope. For any kind of objective it is possible to simulate many different specific processes and systems (\ie, objects); the status of all these objects, \eg, whether they need to be actually in place here and now, whether we evaluate them for future manufacturing, or are making use of them as ideal counterfeit objects, is not controlled by the simulation objective. We will demonstrate this through a case study in Section~\ref{subsec:case-study}.  Nonetheless, requirements for documenting the simulated object, which includes documenting its epistemic status, can for the purpose of metadata standardization be \textit{grouped together with} requirements for documenting the simulation objective. This is under the constraint that we are here only considering the epistemic use of simulation, not the totality of its potential use beyond the creation of knowledge. We demonstrate the usefulness of grouping these two dimensions together, specially for visualization purposes, through the diagram proposed in Section~\ref{subsec:visualizing}.

From the point of view of semiotics, the epistemic status of the object can be, obversely, understood as a feature of the sign-object representation relation, with the model in the role of the sign. A sign-object representation relationship is that between a \textit{sign} (in Peircean semiotics, also called a representamen) and the \textit{object} that this sign \textit{represents}. A simulation is a \textit{semiosis} -- a process within which use is made of the model as a sign~\cite{Horsch21,HCSST21}. Case distinctions relevant to both of these are discussed in recent work by Dur\'an~\cite{Duran23,Duran18}. Concerning the representation relation, \qq{three kinds of models based on their representational capacity emerge, namely, phenomenological models, models of data, and theoretical models}~\cite{Duran18}. Concerning the scientific use of simulation artefacts (including but not limited to the models), in analogy to Steinle's~\cite{Steinle97} analysis of the function of experimental work, Dur\'an~\cite{Duran23} considers simulation work to be \textit{theory-driven} whenever \qq{expectations regarding a simulation output fall within the framework provided by the theory or mathematical model that is implemented by the simulation} or, equivalently, the \qq{design and success of the simulation depends on a given theory}; \ie, paradoxically, the term theory-driven is used exclusively for technical work that operates over a closed epistemic space, merely using theory instead of contributing to its development. Complementing this, Dur\'an~\cite{Duran23} considers \textit{exploratory} (\ie, non-theory-driven) use of simulation technology to comprise \qq{three forms of exploratory simulations: computer simulations as starting points and the continuation of scientific inquiry, as varying parameters, and as scientific prototyping}. While parameter variation is also a technical operation, taking a pre-specified parameter space as given, the other two use-case scenarios considered by Dur\'an~\cite{Duran23} are scientific in nature.





\subsection{Subject matter}
\label{subsec:subject-matter}

The second dimension under which the scope of simulations and simulation artefacts can be considered is subject matter, \ie, \textit{what the simulation is about}. Similar to Section~\ref{subsec:objective-object}, where the purpose of the simulation and the status of the object of reference were analysed as two distinct, but closely connected ways to address this dimension, here two such approaches can be distinguished as well: First, subject matter as a partitioning of logical space~\cite{Yablo14}, and second, describing the topic of a digital artefact by labelling.

There is much debate on what constitutes the subject matter of sentences and, by extension, documents consisting of propositions, or datasets annotated by triples which express propositions. Yablo~\cite{Yablo14} proposes that subject matter essentially is a question, or more properly speaking a partitioning of the space of possible states of affairs, which for practical purposes is equivalent to a question. By this understanding, subject matter is a component of the semantics (as opposed to being just an annotation, or part of the pragmatics). For example, the sentence \qq{DCLXVI is held in Kaiserslautern in 2024} might be equivalently expressed through some predicates $\textsf{ScientificEventPlace}(\textsf{DCLXVI},\, \textsf{Kaiserslautern})$ and $\textsf{ScientificEventYear}(\textsf{DCLXVI}$, \textsf{2024}$)$. As part of its semantics, however, we need to complement this by a description of what the sentence is about, \ie, what question it is answering: First, this could be $\textsf{ScientificEventPlace}(x?,\, \textsf{Kaiserslautern}) \,\land$ \textsf{ScientificEvent\-Year}$(x?,\,\textsf{2024})$, \ie, \qq{what scientific events are being held in 2024 in Kaiserslautern?} Or, second, it could be only about the year of DCLXVI, with the question $\textsf{ScientificEventYear}(\textsf{DCLXVI},\, y?)$ or \qq{what year is DCLXVI being held?} In the second case, the question (or the statement addressing the question) is not \textit{about} the location, and states of affairs with different event locations for DCLXVI would be considered equivalent with respect to the subject matter. Even more generally, we could have asked a third question: \qq{Is DCLXVI held in 2024 or some other year?} The answer would again need to give expression to the fact $\textsf{ScientificEventYear}(\textsf{DCLXVI},\, \textsf{2024})$. In the first and third cases, in contrast to the second one, counterfactual states of affairs where the event is held in 2023 or 2025 would be equivalent to each other, since both are different from 2024, which is what these questions are about. With different subject matters, a sentence will have different meanings even if it asserts the same predicates with the same arguments~\cite{Yablo14}; this is because different questions yield different ways of partitioning the space of possible state of affairs. Previous work based on a case study has established that the approach based on an analysis of subject matter is viable for annotating molecular modelling research artefacts~\cite{HCGKMSSTVS23}.

The above suggests that subject matter can be expressed in a data query language, \eg, as a SPARQL query with which a knowledge base containing propositions on the state of affairs can be queried. The semantics of such a query can be given in the form of a first-order logic formula with free variables; these free variables are the information that is being requested (above, $x?$ and $y?$). The semantics of the answer to the query would be given by two parts, namely, the truth conditions of the propositions \textit{and} the subject matter, \ie, way in which the query partitions the logical space. Such \textit{two-component semantics}, where truth conditions and subject matter are regarded as separate, irreducible components of a proposition's meaning, have recently been developed in detail by Plebani and Spolatore~\cite{PS21} as well as Berto and Hawke~\cite{BH22}.

To others, the topic of a document or a dataset would appear as a collection of labels, \ie, a \textit{bag of words}~\cite{Cichosz23,GN24}. 
A comprehensive catalogue of topic labels suitable for physics-based modelling and simulation was compiled in the Virtual Materials Marketplace project (VIMMP, 2018--2022, H2020 GA no.~760907) through the ontology for training services\footnote{URL: \url{http://molmod.info/semantics/otras.ttl}.} (OTRAS) as part of a larger system of ontologies designed for use by the marketplace platform~\cite{HCSTSLABMGKSFBSC20,HCCS21}. Label catalogues can also be populated automatically by text processing~\cite{ZSAS21}. Once a catalogue is in place, in addition to annotating documents with labels by hand, which is not particularly complicated, a variety of text processing techniques are available for use in assisted or automated annotation~\cite{Cichosz23}; \eg, Gizatullin and Nevzorova~\cite{GN24} evaluate BERTopic~\cite{Grootendorst22}, latent Dirichlet allocation~\cite{AKG17}, and non-negative matrix factorization~\cite{ASN17} for this purpose. Automated annotation becomes necessary when processing large corpora and may also be useful at data harvesting, depending on the degree of interoperability established between the interacting platforms. Naively, it seems easier to adequately determine a topic label, out of a tractable finite number of discrete options, than the subject matter expressed as a research question, of which there are infinitely many; Plebani and Spolatore~\cite{PS21}, however, propose a syntax-based construction for the subject matter of composite expressions based on atomic formulas, which directly translates to a linear-time algorithm. Moreover, it seems plausible that large language models can determine the subject matter to an acceptable degree of accuracy as well, specially if the outcome is subject to human curation, \eg, by a scientific data officer~\cite{SD20}, knowledge management translator~\cite{GSBGGK3LMNPSVW22}, or data steward~\cite{Mons18}.

\section{Simulation artefacts}
\label{sec:simulation-artefacts}

\subsection{Following the European materials modelling community}

The European Materials Modelling Council is an organization with a focus on digitalization in physics-based modelling and simulation. Created out of a European project (EMMC CSA, 2016--2019, H2020 GA no.~723867), it has established itself as an organization with a long-term perspective, funded by its members. Its community includes over 20 presently running EMMC-related projects. Major milestones in the EMMC's work on materials modelling digitalization have been the compilation of the Review of Materials Modelling~\cite{DeBaas17} and, on this basis, the development of a CEN workshop agreement (CWA) reference document for standardized model data (MODA) documentation: CWA 17284:2018 MODA~\cite{Cenelec18}. This documentation standard was made mandatory by the European Commission for a series of projects (mostly, the same as the EMMC-related topics at the time), roughly until 2020. Since then, the EC's expectation of projects' compliance with MODA has been relaxed to some degree, even though it occasionally still appears in policy documents and Horizon call topics, usually alongside CWA 17815:2021 CHADA (characterization data)~\cite{Cenelec21}.

Both MODA and CHADA have been implemented as OWL ontologies (OS\-MO~\cite{HNBC3ELNSSTVC20,HTCSGT21} and CHAMEO~\cite{DGT22}, respectively), elaborating on conceptualizations; but these extended formalisms and improved definitions of concepts have no bearing on the standardization reference documents.
MODA is a metadata standard encompassing simulation workflows and all their aspects at a specified level of detail; namely, a level of detail that is so high that it requires great effort from the people in charge of providing the documentation, but on the other hand, not high enough to double as a script by which simulations can be deployed. We will not further develop the criticism of MODA that has been made elsewhere~\cite{HSP23} and of which its users will be aware; as regards usability, the recent development of Easy-MODA~\cite{KSZVMTDESCPNKKSFGWPSGMLA24}, an environment for generating MODA documentations, promises to improve the situation. Similarly, EMMO ontology development, which is a major body of work undertaken by exactly the same community~\cite{FGGHSFD20,ZMGB23}, but unfortunately not very strongly aligned~\cite{KPHK21} with the same community's work on MODA, will not be discussed here either.

Using MODA, all the aspects of a simulation workflow can be documented; selected concepts related to this are listed in Tab.~\ref{tab:moda-modgra}.

\begin{table}[p!]
   \centering
   \caption{Simulation artefacts (top) and related concepts (bottom) defined in the two CEN workshop agreements CWA 17284:2018~\cite{Cenelec18} and CWA 17960:2022~\cite{Cenelec22}. Neither of the two enables documenting the execution environment or the simulation objective.}
   \bigskip
   {\footnotesize
   \begin{tabularx}{\textwidth}{l | X | X}
   Classification ~ & CWA 17284:2018 MODA ~ & CWA 17960:2022 ModGra ~ \\ \hline
   Model class & \textit{Physics equation:} \qq{equation based on a fundamental physics theory which defines the relations between physics quantities of an entity}. & --- \\
   Model (partial) & \textit{Materials relation:} \qq{materials specific equation providing a value for a parameter}. & --- \\
   \textbf{Complete model} & \textit{Physics-based model:} \qq{solvable set of one physics equation and one or more materials relations}. & \textit{Model:} \qq{mathematical predictive representation of something}. \\
   Numerics & \textit{Computational representation}: Variables and equations chosen to implement the model numerically. & \textit{Surrogate object:} \qq{substitute of a (composite) object/entity mimicking the substituted behaviour}. \\
   & \textit{Solver:} \qq{techniques used to numerically solve a particular physics-based model}. & \textit{Control capacity:} Information item that can influence the control flow. \\
   Simulation software & \textit{Software tool} (field 3.2). & --- \\ 
   \textbf{Simulation input} & \textit{Simulated input} (field 2.5). & Represented by tokens. \\
   Execution environment & --- & --- \\
   \textbf{Simulation} & \textit{Workflow:} \qq{A graphical representation of a simulation can be given in a diagram and is called workflow}. & \textit{Process model topology:} Generalized Petri net representing both the process and the process model. \\
   \textbf{Simulation output} &  \textit{Processed output} (field 4.1). & Represented by tokens. \\
   \textbf{Knowledge claim} & \textit{Post-processing}: \qq{operations on raw output of solvers}. &  \\ \hline
   Simulated object & \textit{Entity:} \qq{four types of entity:} Electron entity, atom entity, mesoscopic entity, and continuum volume entity. & \textit{Physical object:} \qq{object/entity that either does, or at least, could exist in the physical world}. \\
   & \textit{Quantity}: \qq{property of a phenomenon, body or substance, where the property has a magnitude that can be expressed as a number and a reference}. & \textit{Physical capacity:} Conserved quantity (energy, matter, momentum, charge) or balanced quantity (not necessarily conserved, \eg, enthalpy). \\
   & \textit{Physical system state:} \qq{values of the physics quantities [\dots] at an instant of time}. & \textit{Physical process:} \qq{transforms a physical object [\dots] into another}. \\
   Simulation objective & --- & --- \\
   Research question & \textit{User case aspect} (field 1.1). & --- \\ \hline
   \end{tabularx} }
   \label{tab:moda-modgra}
\end{table}

\subsection{Following the computer aided process engineering community}

ModGra~\cite{Cenelec22} (\qq{model graphs}) and DEXPI Process~\cite{COTHT24} (where DEXPI stands for \qq{data exchange in the process industry}) are conventions for denoting physicochemical processes and models of such processes; both have recently been proposed within the computer aided process engineering community. Aspects shared by the approaches include that they are data models and graphical notations at the same time, and that in the graphical representation, depending on context, arrows represent flow of information or exchange of matter or energy. We will here restrict ourselves to analysing ModGra, which has reached the higher degree of maturity by going through a community consultation and formalization process that was moderated by the Italian standardization organization UNI.

In ModGra, the description of a model is given the form of a description of the modelled system; thereby, the documentation of the physical and the virtual can be made consistent by design. ModGra is formalized through CWA 17960:2022~\cite{Cenelec22}; however, the approach underlying it had already been established by Preisig~\etal~\cite{EP19,Preisig21a,Preisig22,PHFKK21} in the years preceding its standardization. The main structure described by the CWA ModGra is the process model topology, a generalized Petri net which, like a conventional Petri net~\cite{Petri62}, consists of places, transitions, arcs, and tokens which can be assigned to the places and consumed or created by the transitions. Much of ModGra is obscured by idiosyncratic terminology and notation, up to what only can be mistakes in the CWA 17960 reference document.\footnote{In the CEN-issued reference document~\cite{Cenelec22}, the definition \qq{3.9 Interface -- Interface is the transfer of state information} is immediately followed by \qq{3.10 Intraface -- Intraface is the Transfer of state information.} So interface and intraface are the same, except for the capitalization of the word \qq{transfer}. From Preisig~\cite{Preisig21a}, however, it would seem that an \qq{intraface} is a conventional Petri net transition, whereas an \qq{interface} is like a transition, except that it does not create or consume any tokens. Another ungrammatical and nonsensical definition from the CWA document is that of \qq{3.18 Tokens: Control capacities,} which is \qq{Tokens are information bits place.}} For our purposes, in any case, the choice of notational elements is less relevant than what the elements are stated to represent.

Core concepts from ModGra are collated with those from MODA in Tab.~\ref{tab:moda-modgra}.

%

\begin{figure}[p!]
    \centering
    \caption{Object-objective abstractness diagram: A proposed visualization technique for the landscape of physics-based simulation use cases according to two dimensions of scope: The abstractness of the object of reference and the abstractness of the intention. The ellipses are examples for how the diagram could be used to contrast typical use cases with each other. (It is not meant as a comprehensive catalogue of use cases.)}
    \bigskip
    \includegraphics[width=\textwidth]{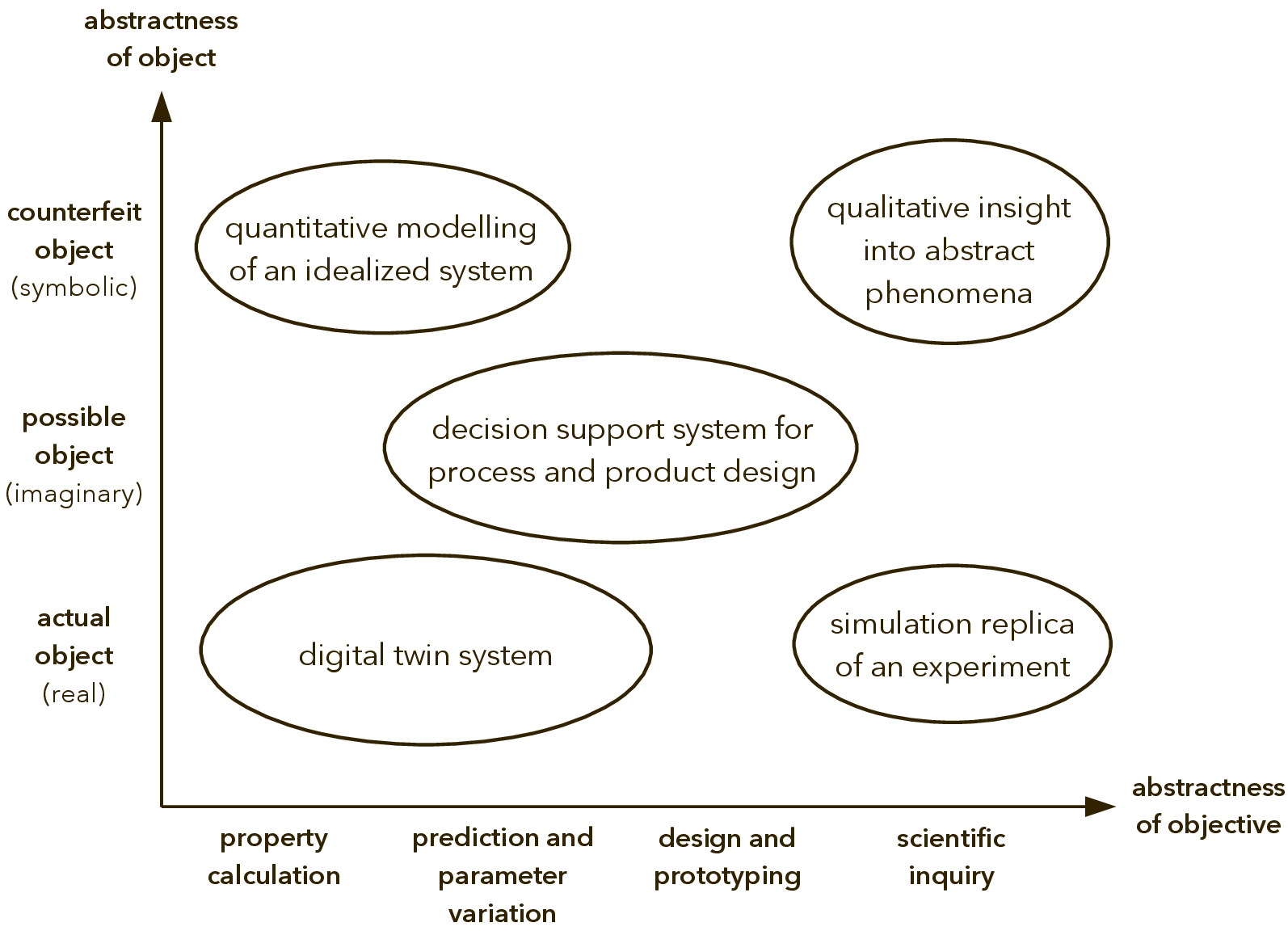}
    \label{fig:landscape}
\end{figure}

\section{Conceptual landscape}
\label{sec:landscape}

\subsection{Visualizing the landscape of the scope of simulations}
\label{subsec:visualizing}

Based on the discussion in Section~\ref{subsec:objective-object}, we propose the \textit{object-objective abstractness diagram} (\cf~Fig.~\ref{fig:landscape}) as a tool for positioning diverse simulation scenarios on a two-dimensional landscape. The horizontal axis positions use cases to the left if they are \qq{theory-driven} following Dur\'an~\cite{Duran23}, and more to the right if they make use of \qq{exploratory strategies}; recall that Dur\'an~\cite{Duran23} slightly counter\-intuitively refers to simulations as theory-driven if they do \textit{not} engage in theory development. We can similarly say that simulation-driven practices are placed on the left if they are technical and operate over a closed epistemic space, and right if they are scientific and operate over an open epistemic space~\cite{Tulatz18}. The vertical axis represents the degree of idealization inherent in the simulated object: For example, a digital twin requires an actual object as its physical twin~\cite{ZLK22}, while process and product design operates over an object that is, as of now, not in place, but the construction of which is at least considered possible. Highly idealized models such as, \eg, the hard-sphere fluid or the Lennard-Jones (LJ) fluid, \textit{can} be used as models of real systems, in which case the use case belongs on the lower side of the diagram. A typical use of such models, however, is as a benchmark system against which physical theories can be validated. Lenhard~\etal~\cite{LSH24a} summarize: \qq{The LJ \textit{fluid} is fundamentally different. It is \textit{defined} as the fluid whose molecular interactions exactly follow the LJ potential. In other words, it is not a more-or-less adequate stepping-stone toward finding the true potential of some real fluid, but postulates a fluid that does not exist in reality.} In such cases, the model represents an idealized counterfeit object, \ie, an object that is more perfect than it would be possible to achieve in a physical experiment, and therefore the best possible object against which a theory can be validated.

\begin{table}[p!]
   \centering
   \caption{Overview over the research articles included in the case study from Section~\ref{subsec:case-study}, surveying work from the Thermodynamics Group at TU Berlin following the approach to representing the landscape of simulation scope introduced in Section~\ref{subsec:visualizing}.}
    \bigskip
   {\footnotesize
   \begin{tabularx}{\textwidth}{l l | X | X}
   Research work ~ & & Object of reference ~ & Simulation objective ~ \\ \hline
   Abbas \& Vrabec~\cite{AV21} & & dual-loop organic Rankine cycle (ORC), varying working fluids and temperature & designing the ORC optimally, \ie, choosing the best working fluids \\ \hline
   Chatwell~\etal~\cite{CGGSV21} & & binary mixtures of carbon dioxide and ethanol close to critical conditions (specifically, $p$ = 10 MPa constant) & qualitative understanding of phenomena at the Widom line; quantitative description of the binary system by simulation and experiment \\ \hline
   Fingerhut~\etal~\cite{FGNSMLPCBSKHV21} & Fig.~1 & Mie and TT potentials & measure parallel speedup \\
   & Fig.~3 & quaternary LJ mixture & thermodynamic factor matrix for diffusion coefficients \\
   & Fig.~7 & binary LJ \textit{as} liquid Ar+Kr & linear transport coefficients \\
   & Fig.~10 & N$_2$ + O$_2$ mixtures & vapour-liquid equilibria, comparing simulation methods to each other and to experimental data \\ \hline
   Guevara~\etal~\cite{GFV21} & & mixtures containing any 2, 3, or 4 of: Water, methanol, ethanol, and isopropanol & predict and measure properties, validate predictive power of molecular models \\ \hline
   Heinen~\etal~\cite{HHDSLV22} & & LJTS fluid \textit{as} Ar: Droplets undergoing coalescence & qualitative theoretical insights and comparing MD to continuum modelling \\ \hline
   Homes~\etal~\cite{HHVF21} & & LJTS liquid evaporating into a vacuum; LJTS considered as a reference fluid & qualitative analysis of evaporation phenomena, quantitative description of LJTS \\ \hline
   Nitzke~\etal~\cite{NSSPGV23} & & fuel injection in combustion engines; also, \{CH$_4$, C$_6$H$_6$\} + \{N$_2$, O$_2$\} binary mixtures & qualitative understanding of the process, comparison of modelling approaches \\ \hline
   R\"o\ss{}ler~\etal~\cite{RASV22} & & 29 pure fluids; specificially, the Joule-Thomson inversion curves of these fluids & compare empirical and molecular EOS to molecular simulation in order to validate the molecular models \\ \hline
   \v{S}ari{\'c}~\etal~\cite{SGV22} & & binary mixtures including supercritical CO$_2$ as a solvent; six different solute components are considered & phenomena at the Widom line; comparison of molecular models to reference correlations and EOS \\ \hline
   \end{tabularx} }
   \label{tab:case-study}
\end{table}

\subsection{Case study}
\label{subsec:case-study}

To survey the landscape for the case of representative work done by a concrete research group, nine research articles from the Thermodynamics Group at TU Berlin are evaluated as regards their scope, specifically, the \textit{abstractness of the objective} and the \textit{abstractness of the object}, \cf~Fig.~\ref{fig:landscape}. To obtain a sample representative of successful research, among all the papers published 2021 or later with a corresponding author from the research group, excluding any exclusively experimental research work, the nine papers with the largest number of citations according to ISI Web of Knowledge\footnote{Date of evaluation: 13th November 2024.} were selected for inclusion in the case study.

Most of these works are straightforward to position on the landscape~\cite{AV21,CGGSV21,GFV21,HHVF21,RASV22,SGV22}, \cf~Tab.~\ref{tab:case-study}. In addition, the work by Fingerhut~\etal~\cite{FGNSMLPCBSKHV21} addresses a whole spectrum of use cases. Mainly, it presents version 4.0 of the molecular dynamics~(MD) and Monte Carlo code \textit{ms2}; at this occasion, a variety of use cases are covered that can benefit from the new features available in that version of their code.\footnote{URL: \url{https://ms-2.de/}.} The modal status of the objects of reference considered in this use cases spans a broad spectrum: From the Mie and Tang-Toennies~(TT) potentials~\cite[Fig.~1]{FGNSMLPCBSKHV21} and a mixture of four kinds of LJ fluids~\cite[Fig.~3]{FGNSMLPCBSKHV21}, which are investigated as pure counterfeit objects, over an argon-krypton mixture~\cite[Fig.~7]{FGNSMLPCBSKHV21}, to the comparison of experimental and simulation data for the vapour-liquid equilibrium of the two main components of air~\cite[Fig.~10]{FGNSMLPCBSKHV21}; objectives include performance measurements to evaluate the numerics as well as the computation of thermodynamic and transport properties. There, while the Ar+Kr system nominally represents a really possible system, it is arguably considered \textit{because} both components are modelled as LJ fluids; the same can be said about the LJ truncated-stifted (LJTS) \qq{argon} system considered by Heinen~\etal~\cite{HHDSLV22}.

The case of Nitzke~\etal~\cite{NSSPGV23} is of interest in that it, as its eventual technical use cases, considers injection of fuel and mixing of the fuel with oxygen (in the case of rockets) or air (in the case of cars). However, the actual simulations are done for binary mixtures of, first, methane \textit{or} cyclohexane with, second, oxygen \textit{or} nitrogen, \ie, for four kinds of binary systems that can and do exist in reality and for which there are experimental results, but which are greatly simplified in comparison to the multicomponent mixtures present in the technical use cases. Accordingly, pure methane or pure cyclohexane, respectively, is said to represent the fuel. Nitrogen is considered not because it represents air, but because it \qq{is often used by experimentalists as non-oxidizing substitute for oxygen}~\cite{NSSPGV23}.

\begin{figure}[p!]
    \centering
    \caption{Object-objective abstractness diagram for the case study (see also Tab.~\ref{tab:case-study}).}
    \bigskip
    \includegraphics[width=\textwidth]{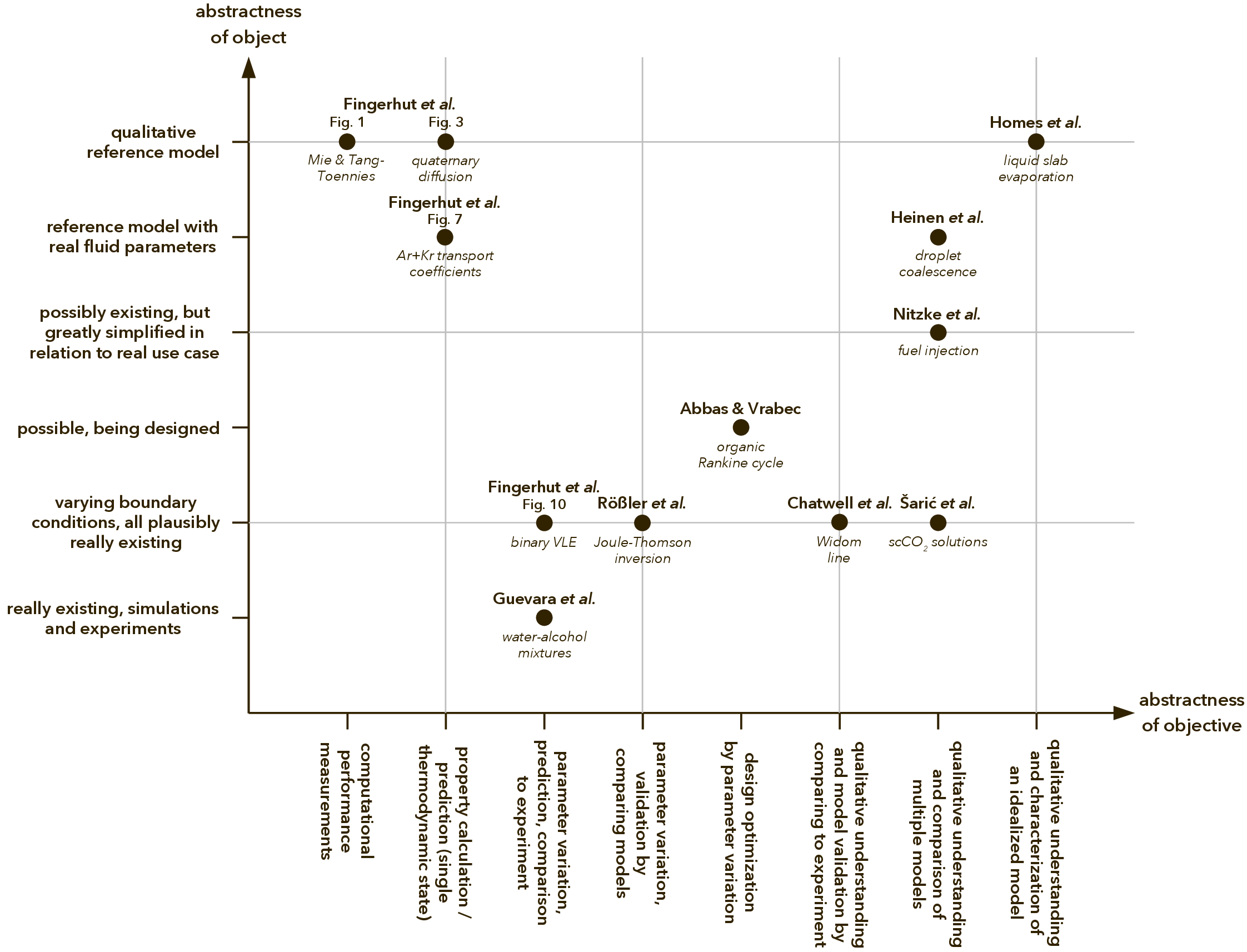}
    \label{fig:case-study}
\end{figure}

Based on this, to illustrate how the object-objective abstractness diagram would be used in practice, we can order the elements of our sample by abstractness of the objective $x$
\begin{align}
   \textnormal{Fingerhut~\etal~\cite{FGNSMLPCBSKHV21} Fig.~1} \quad<_x\quad & \textnormal{Fingerhut~\etal~\cite{FGNSMLPCBSKHV21} Fig.~3, Fig.~7} \nonumber \\
   \quad<_x\quad & \textnormal{Fingerhut~\etal~\cite{FGNSMLPCBSKHV21} Fig.~10,} \nonumber \\
   & \textnormal{Guevara~\etal~\cite{GFV21}} \nonumber \\
  ·  \quad\dots\quad & \nonumber \\
   \quad<_x\quad & \textnormal{Heinen~\etal~\cite{HHDSLV22}, Nitzke~\etal~\cite{NSSPGV23},} \nonumber \\
   & \textnormal{\v{S}ari\'c~\etal~\cite{SGV22}} \nonumber \\
   \quad<_x\quad & \textnormal{Homes~\etal~\cite{HHVF21},}
\end{align}
and by abstractness of the object $y$
\begin{align}
   \textnormal{Guevara~\etal~\cite{GFV21}} \quad<_y\quad & \textnormal{Chatwell~\etal~\cite{CGGSV21}, Fingerhut~\etal~\cite{FGNSMLPCBSKHV21} Fig.~10,} \nonumber \\
   & \textnormal{R\"o\ss{}ler~\etal~\cite{RASV22}, \v{S}ari\'c~\etal~\cite{SGV22}} \nonumber \\
   \quad\dots\quad & \nonumber \\
   \quad<_y\quad & \textnormal{Fingerhut~\etal~\cite{FGNSMLPCBSKHV21} Fig.~7, Heinen~\etal~\cite{HHDSLV22}} \nonumber \\
   \quad<_y\quad & \textnormal{Fingerhut~\etal~\cite{FGNSMLPCBSKHV21} Fig.~1, Fig.~3,} \nonumber \\
   & \textnormal{Homes~\etal~\cite{HHVF21}.}
\end{align}
This results in the diagram shown in Fig.~\ref{fig:case-study}. The case study illustrates how the object-objective abstractness diagram can be tailored to the landscape of simulation-based science and engineering considered in a particular context. The distribution of the works included in the case study across the landscape also shows that the status of the modelled object and the kind of work done with the model, in terms of what kind of knowledge is being pursued, are independent dimensions. They must be accounted for separately and, consequently, can be used as a landscape visualization technique for the considered domain of research.

\subsection{Simulation artefact documentation requirements}

\begin{figure}[p!]
    \centering
    \caption{Left: E-R diagram with relations (diamonds) that should be present in an ontology for documenting the scope of simulation artefacts. Right: Suggested knowledge graph shape for data ingest of a simulation object; there, redundant relations are eliminated. The symbol \textsf{R} is used for relations that are subproperties of the Peircean representation relation~\cite{Horsch21,HCCS21,HS22,Peirce55} with the meaning \qq{represents} or \qq{is a representamen (sign) for.} Arrows designate \textit{to-one} relations, and double lines designate \textsf{owl:someValuesFrom} restrictions; \eg, the relation between simulation and simulation output is \textit{one-to-one}: For every simulation, there is one simulation output object (the I/O data, respectively, are here understood to be grouped into single objects), and vice versa. The relation between simulation input and simulation is \textit{$n$-to-one}, with each simulation requiring exactly one simulation input object, while multiple simulations can be conducted using the same input. Numbers inside circles: See Tab.~\ref{tab:implementation}.}
    \bigskip
    \includegraphics[width=\textwidth]{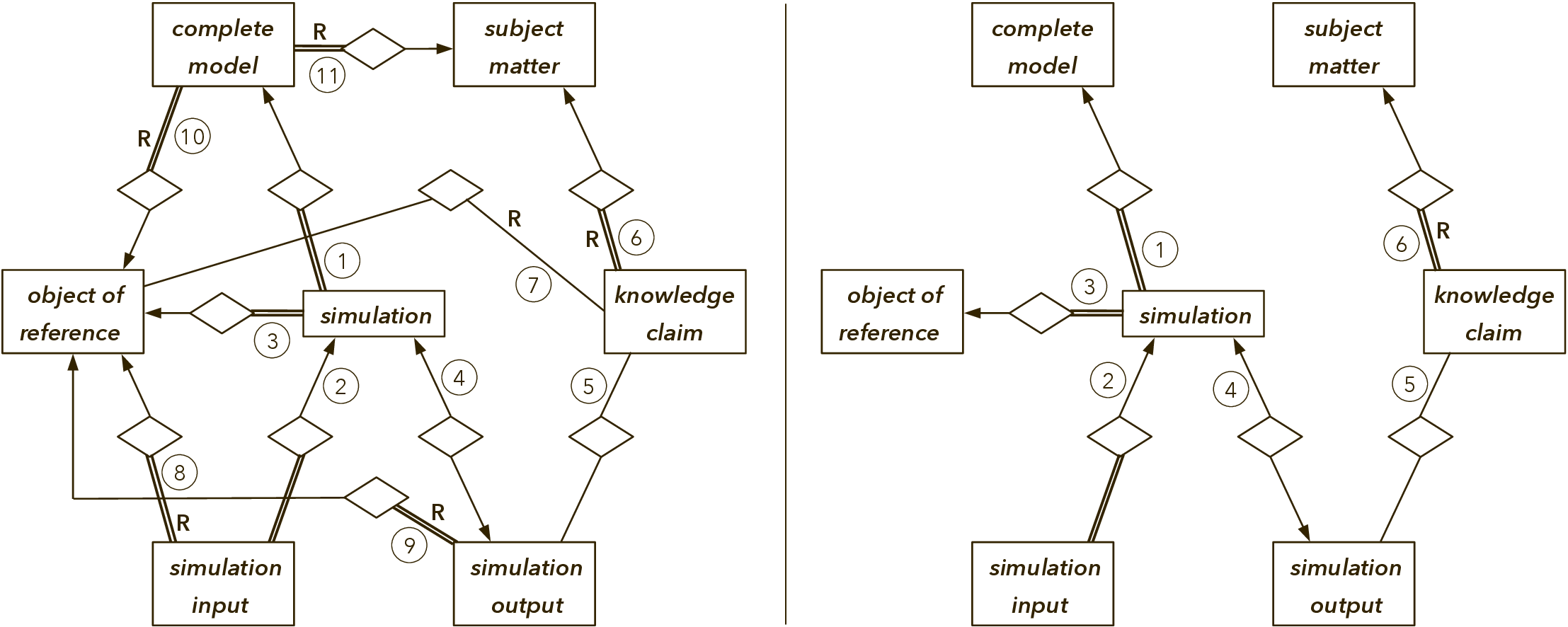}
    \label{fig:simulation}
\end{figure}

We now integrate the landscape explored above into a tentative schema, intended as input for future discussions on epistemic metadata standardization for simulation artefact documentation.
For this purpose, we formulate requirements for metadata standardization on the scope of simulation artefacts based on the juxtaposition from Tab.~\ref{tab:moda-modgra}, focusing on simulation artefacts that can be documented using both MODA and ModGra; these are the entries that are shown in bold face in the left column of Tab.~\ref{tab:moda-modgra}. This is extended by the addressed subject matter (Section~\ref{subsec:subject-matter}) and the simulated object (Section~\ref{subsec:objective-object}), \ie, the object of reference, taking the role of the \textit{object} within the framework of Peircean semiotics~\cite{Horsch21,HCCS21,HS22,Peirce55}. The subject matter corresponds to the entry \qq{research question} in Tab.~\ref{tab:moda-modgra} and can in principle be documented in MODA -- so far, however, only as a free-text entry without any further semantics. The E-R diagrams on the left side of Fig.~\ref{fig:simulation} contains concepts and relations that should be present in the ontology. These recommendations can be taken into consideration for the ongoing refactoring of the PIMS-II mid-level ontology of cognitive processes\footnote{PIMS-II: Physicalistic interpretation of modelling and simulation -- interoperability infrastructure. URL: \url{http://www.molmod.info/semantics/pims-ii.ttl}.}~\cite{Horsch21,HCGKMSSTVS23,HS22} into the MSO-EM system of mid-level ontologies~\cite{HCTCDKSSTSKAS24,HRVBCGJKLSSVWSTCA24},\footnote{MSO-EM: Ontologies for modelling, simulation, optimization, and epistemic metadata. Persistent URL: \url{https://www.purl.org/mso-em}. Public development repository: \url{https://github.com/HE-BatCAT/mso-em}.} aligned with the foundational ontology DOLCE~\cite{BM10,PG15}. In addition, the simulation requires an agent and an intention, \ie, an objective~(Section~\ref{subsec:objective-object}); these requirements, which are not visualized in Fig.~\ref{fig:simulation}, can be realized through the MSO-EM agency module\footnote{URL: \url{https://batcat.info/semantics/mso-em/agency.ttl}.} which is based on Conte's taxonomy of agents~\cite{Conte12}.

During data ingest for a simulation object, it would be redundant to use all these relations at once; \eg, the subject matter of the complete model is the addressed research question, and the subject matter of the obtained knowledge claim(s) is again the same research question. Similarly, the complete model, the simulation input, and the simulation output are signs for the same object (\ie, the simulated system), which also fills the \textit{object} role of the simulation as a Peircean semiosis. The right-hand side of Fig.~\ref{fig:simulation} visualizes a non-redundant knowledge graph shape that could be used as part of a data exchange interface specification, \eg, using SHACL (shapes constraint language) or OO-LD.\footnote{OO-LD: Object-oriented linked data. URL: \url{https://github.com/OO-LD}.}

\subsection{Implementation}

The implementation of the simulation artefact documentation is done within the framework of the MSO-EM ontologies~\cite{HCTCDKSSTSKAS24,HRVBCGJKLSSVWSTCA24}. These ontologies are aligned with DOLCE~\cite{BM10,PG15}, which follows the paradigm of endurantism. Accordingly, a purposeful action (such as a simulation) is classified as a perdurant. It is carried out by a goal-directed agent (a physical endurant) with an intention, which is conceptualized as a quality inherent in the goal-directed agent, and therefore, since the agent is a physical endurant, as a physical quality, as required by DOLCE. As an endurant, the agent does not have any temporal parts; instead, the value of the quality can vary over time. Since within DOLCE, physical endurants have physical qualities, while perdurants have temporal qualities, it follows that the intention (a physical quality) cannot be a quality inherent in the action. Instead, the relation between the purposeful action and the intention must be classified as a mediated relation, since it requires a third entity, the agent. As the value of the agent's intention with respect to the action varies over time, temporal parts of the action are associated with different intention values. It therefore seems reasonable to accept that a single purposeful action can be associated with multiple intention values;  within the framework of DOLCE, by means of a mediated relation. If, instead, we needed the intention associated with the action to be a single entity, the intention for the temporal whole might be the mereological sum of the intentions associated with the temporal parts. This could be covered by the theory of collective entites proposed by Masolo~\etal~\cite{MVFBP20}. The relation between the agent and the intention value (in DOLCE, a quale) is here subsumed under the DOLCE relation \textsf{exact-location}, where as the relation between the action and the agent's intention is subsumed under \textsf{generic-location}.

\begin{table}[p!]
   \centering
   \caption{Implementation of the relations from Fig.~\ref{fig:simulation} in the MSO-EM system of mid-level ontologies; in parentheses: Alignment with the foundational ontology DOLCE. The symbol $\top$ here stands for \textsf{owl:Thing}.}
    \bigskip
   {\footnotesize
   \begin{tabular}{r l | l | l}
   no. ~ & relation & domain & range \\ \hline
   1 ~ & \textsf{evaluates model} & \textsf{simulation} (\textsf{event}) & $\top$ \\
     ~ & (\textsf{specific constant dependent}) & & \\ \hline
   2 ~ & \textsf{involves sign} (\textsf{participant}) & \textsf{semiosis} (\textsf{event}) & $\top$ \\ \hline
   3 ~ & \textsf{involves referent} & \textsf{cognition} (\textsf{event}) & $\top$ \\
     ~ & (\textsf{weak connection}) & & \\ \hline
   4 ~ & \textsf{involves interpretant} & \textsf{semiosis} (\textsf{event}) & $\top$ \\
     ~ & (\textsf{participant}) & & \\ \hline
   5 ~ & \textsf{is based on} & \textsf{claim} (\textsf{proposition}) & \textsf{articulation} \\
     ~ & (\textsf{generic constituent}) & & (\textsf{non-physical endurant}) \\ \hline
   6 ~ & \textsf{has subject matter} & (\textsf{proposition}) & (\textsf{abstract}) \\
     ~ & (\textsf{generically dependent on}) & & \\ \hline
   7 ~ & \textsf{is about} & $\top$ & $\top$ \\
     ~ & (\textsf{generically dependent on}) & & \\ \hline
   8 ~ & \textsf{represents} & $\top$ & $\top$ \\
     ~ & (\textsf{generically dependent on}) & & \\ \hline
   9 ~ & \textsf{represents} & $\top$ & $\top$ \\
     ~ & (\textsf{generically dependent on}) & & \\ \hline
   10 ~ & \textsf{articulates model of} & \textsf{articulation} & $\top$ \\
      ~ & (\textsf{generically dependent on}) & (\textsf{non-physical endurant}) & \\ \hline
   11 ~ & \textsf{is about} & $\top$ & $\top$ \\
      ~ & (\textsf{generically dependent on}) & & \\ \hline
   \end{tabular} }
   \label{tab:implementation}
\end{table}

In addition, the relations from the MSO-EM mid-level ontologies that can be used to document the pattern from Fig.~\ref{fig:simulation} are specified in Tab.~\ref{tab:implementation}; relations and their domains and ranges are shown jointly with the most specific relations and concepts from DOLCE, respectively, under which they are subsumed.

\section{Conclusion}
\label{sec:conclusion}

The main results from this work are, first, the \textit{object-objective abstractness diagram} as a tool for analysing the conceptual landscape and, second, the analysis of minimally required concepts and relations for epistemic metadata ontology development regarding the scope of simulation artefacts. We propose to focus on documenting, beside the \textit{simulation} instance itself, three kinds of simulation artefacts that are supported by both MODA and ModGra: The \textit{complete model}, \ie, an articulation of a completely determined and solvable set of equations including all parameters and boundary conditions, the \textit{simulation input}, and the \textit{simulation output}; in addition to these three kinds of simulation artefacts, the \textit{knowledge claims} made by a researcher upon analysing the simulation output need to be included as a core element of the epistemic metadata. We propose to include the \textit{object}, the \textit{objective}, and the \textit{subject matter} among the metadata for the simulation, and for the subject matter, we recommend a formalization based on the \textit{research question}, rather than topic labelling based on the bag-of-words approach. In this way, two-component semantics separating truth conditions and subject matter can be supported, \ie, it can be taken into account that part of the meaning of a research result is contained in the question that was being addressed. These results are offered to the Knowledge Graph Alliance and its community as a contribution to the process and collective effort of agreeing on the next generation of good practices, which will make simulation artefacts XAIR and simulation workflows explainable.

This process, however, will only be successful to the extent that it engages diverse groups of people who can assess and support the work each from their own perspective, and eventually facilitate the uptake of the results. We hope that the DCLXVI 2024 International Workshop will meet the stakeholders' expectations of such a process, and appeal to all to contribute to its further development.

\bigskip

\noindent
\textit{Acknowledgment.\/} The co-authors F.A.M.~and M.T.H.~acknowledge funding from the EU's Horizon Europe research and innovation programme under grant agreements 101137725~(BatCAT) and 101138510~(DigiPass CSA). The co-author J.V.~acknowledges financial support by Deutsche Forschungsgemeinschaft~(DFG) under grant VR~6/16.

\bibliographystyle{klunummod}
\bibliography{dclxvi-physics-based-scope}
\end{document}